\documentclass[epj]{svjour}
\usepackage{graphicx}
\usepackage{subfigure}
\usepackage{amssymb}
\usepackage{amsmath}
\usepackage{amsfonts}
\usepackage{cancel}
\usepackage{wasysym}
\usepackage{lmodern,dsfont}
\usepackage{ulem}
\usepackage[usenames]{color}

\newcommand{\be}{\begin{equation}}
\newcommand{\ee}{\end{equation}}
\newcommand{\ba}{\begin{eqnarray}}
\newcommand{\ea}{\end{eqnarray}}
\newcommand{\non}{\nonumber \\[2mm]}

\begin{document}

\title{Unitarized Chiral Perturbation Theory 
in a finite volume:\\ scalar meson sector}

\author{M.~D\"oring\inst{1}\and U.-G.~Mei\ss ner\inst{1,2}\and
E.~Oset\inst{3}\and A.~Rusetsky\inst{1}}

\institute{
Helmholtz-Institut f\"ur Strahlen- und Kernphysik (Theorie) and Bethe Center
for Theoretical Physics,  Universit\"at Bonn,\\ Nu\ss allee 14-16, D-53115
Bonn, Germany
\and
Forschungszentrum J\"ulich, J\"ulich Center for Hadron Physics, Institut f\"ur
Kernphysik  (IKP-3) and Institute for Advanced Simulation (IAS-4), D-52425
J\"ulich, Germany
\and
Departamento de F\'{\i}sica Te\'orica and IFIC, Centro Mixto Universidad de
Valencia-CSIC, Institutos de Investigaci\'on de Paterna, Aptdo. 22085, 46071
Valencia, Spain
}

\abstract{
We develop a scheme for the extraction of the properties of the scalar mesons
$f_0(600)$, $f_0(980)$, and $a_0(980)$ from lattice QCD data.  This scheme is
based on a two-channel  chiral unitary approach with fully relativistic
propagators in a finite volume. In order to discuss the feasibility of finding
the mass and width of the scalar resonances, we analyze synthetic lattice data
with a fixed error assigned, and show that the framework can be indeed used for
an accurate determination of resonance pole positions in the multi-channel
scattering.
}

\maketitle

\section{Introduction}
\label{Intro}
The application of lattice QCD techniques to obtain ha\-dro\-nic spectra and
resonances is  catching up and looks very promising for the near
future~\cite{Nakahara:1999vy,sasaki,Mathur:2006bs,Basak:2007kj,Bulava:2010yg,Dudek:2011tt,Morningstar:2010ae,Foley:2010te,Baron:2009wt,Alford:2000mm,Kunihiro:2003yj,Suganuma:2005ds,Suganuma2,Suganuma3,Hart:2006ps,Hart2,Wada:2007cp,Prelovsek:2010gm,Prelovsek2,Prelovsek:2010kg}. However, as it is
well known, resonances do not correspond to isolated energy levels in the
(discrete)  spectrum of the QCD Hamiltonian measured on the lattice,  and an
additional effort is needed to extract the parameters of a resonance  (its mass
and width) from the ``raw'' lattice data. In case of elastic  scattering,
the pertinent procedure is well known under the name of L\"uscher
framework~\cite{luscher,Luscher:1990ux}. In this framework, for a system
described by a given  quantum-mechanical Hamiltonian, one relates the measured
discrete value of the energy  in a finite volume to the scattering phase shift
at the same energy, for the same system in the infinite volume. Consequently,
studying the volume-dependence of the discrete spectrum of the lattice QCD
gives the energy dependence of the elastic scattering phase shift and
eventually enables one to locate the resonance pole positions.

Recently, the L\"uscher approach has been generalized to the case of
multi-channel scattering.  This was done in Refs.~\cite{Liu:2005kr} on the
basis of potential scattering theory, while the
Refs.~\cite{Lage:2009zv,akaki} use non-relativistic effective field theory
(EFT) for this purpose. As discussed in Ref.~\cite{akaki}, there is a
fundamental difference between L\"uscher approaches in the elastic and the
inelastic cases: whereas in the former, one aims at the extraction of a single
quantity (the scattering phase shift) from the single measurement of the energy
level, there will be, e.g., three different observables at a single energy in
case of two-channel scattering (conventionally, the two-channel $S$-matrix
is parameterized by two scattering phases and the inelasticity parameter). In
order to circumvent the above problem, in Ref.~\cite{akaki} it has been
proposed to impose twisted boundary conditions (b.c.). Performing the
measurements of the spectrum at different values of the twisting angle provides
one with data, which are in principle  sufficient to determine all
$S$-matrix elements in the multi-channel scattering independently. Alternative
proposals, e.g., using asymmetric lattices, will be briefly considered in the
present paper. 

In this paper we consider a framework for the extraction of the scalar
resonance parameters, based on unitarized Chiral Perturbation Theory (UCHPT).
In the infinite volume, this model is very  successful, and reproduces well the
$\pi \pi/\pi\eta$ and $ K \bar{K}$ data up to 1200 MeV. The resonances
$f_0(600),f_0(980),a_0(980)$ are generated dynamically  from the coupled
channel interaction \cite{npa,Pelaez:2003dy,Kaiser:1998fi,Locher:1997gr}. In
this paper, we consider the same model in a finite volume to produce the
volume-dependent  discrete energy spectrum. Reversing the argument, one may fit
the parameters of the chiral potential to the measured energy spectrum on the
lattice and, at the next step, determine the resonance locations by solving the
scattering equations in the infinite volume.

In this work we address two main issues. The first one is the use of fully
relativistic propagators in the effective field theory framework in a finite
volume. It is demonstrated that this allows one to avoid the sub-threshold
singularities which are inherent to the L\"uscher approach. In case of the
multi-channel scattering, these sub-threshold singularities may come
dangerously close to the physical region or even sneak into it. We also study
the numerical effect of the use of relativistic propagators on the
finite-volume spectra and on the extraction of the physical observables from
the lattice data.

Second, we discuss in detail the analysis of ``raw'' lattice data for the
multi-channel scattering. To this end, we supplement lattice data by a piece of
the well-established prior  phenomenological knowledge that stems from
UCHPT, in order to facilitate the extraction of the  resonance parameters. In
particular, it will be shown that, with such prior input, e.g., the extraction
of the pole position from the data corresponding only to the periodic b.c., is
indeed possible. 
 
In the present paper, in order to verify the above statements, we shall analyze
``synthetic'' lattice data. To this end, we produce  energy levels by using
UCHPT in a finite volume, assume  Gaussian errors for each data point, and then
consider these as the lattice data, forgetting how they were produced  (e.g.,
forgetting the parameters of the effective chiral potential and the value of
the cutoff).  In the analysis of such synthetic data, we shall test our
approach, trying to establish resonance masses and widths from the fit to the
data. Note that our procedure is similar to the one used previously in
Ref.~\cite{Bernard:2008ax} for the case of the $\Delta$-resonance in $\pi
N$ elastic scattering. Recently, the problem of extracting continuum 
quantities from synthetic lattice data has also been also discussed in
Ref.~\cite{Torres:2011pr} in the context of meson-meson interaction with charm
and the $D_{s*0}(2317)$ resonance. The extraction of continuum quantities, 
at higher energies, may also
be performed using dynamical coupled-channels approaches as shown in 
Ref.~\cite{Doring:2011ip}.

The paper is organized as follows. In Sec.~\ref{sec:formalism} we develop
the relativistic effective field theory formalism in a finite volume. In
section~\ref{sec:twisted-asym} we elaborate our method in coupled channels and
discuss the use of twisted b.c. and asymmetric lattices for the extraction of
the multi-channel resonances.  Finally, in section~\ref{sec:fit} we give a
detailed discussion  of the fit by using UCHPT language in a finite volume.
Section~\ref{sec:concl} contains our conclusions.

\section{Formalism}
\label{sec:formalism}

\subsection{General setting}
\label{sec:2a}

Below, we briefly present our approach to the finite volume problem within the
setting of UCHPT. Unless stated otherwise,  we shall restrict ourselves to
the case with  total isospin $I=0$ (the case with $I=1$ can be considered
analogously and is briefly discussed  in section~\ref{subsec:a0}). We thus
start by using the chiral unitary approach in two channels,  $\pi \pi$ and $ K
\bar{K}$, with the potential obtained from the lowest order chiral Lagrangians
\cite{gasser}. We follow the approach of Ref.~\cite{npa} using the coupled 
channel Bethe-Salpeter (BS) equation. As shown in Refs.~\cite{npa,angels}
using  a certain renormalization procedure, or in Refs.~\cite{nsd,ollerulf}
using dispersion relations, in a certain partial wave one can write the BS
equations  in a factorized form, where the on shell potential and scattering 
amplitudes are factorized out of the $VGT$ integral term of the  BS equations
and, as a consequence, the integral coupled equations become simple algebraic
equations.  These BS equations  are written as 
\cite{npa}
\be
T=[1-VG]^{-1}V
\label{bse}
\ee
where $V$ is a $2\times 2$ matrix accounting for the S-wave $\pi \pi \to \pi
\pi$, $K \bar{K} \to K \bar{K}$ and  $\pi \pi \to K \bar{K}$ potentials and 
$G={\rm diag}\,(G_1,G_2)$  is a diagonal matrix that accounts for the loop
function of the two-meson  propagators in the intermediate states, given by
\ba
G_j&=&\int\limits^{|\vec q|<q_{\rm max}}
\frac{d^3\vec q}{(2\pi)^3}\frac{1}{2\omega_1(\vec q)\,\omega_2(\vec q)}
\non &\times&
\frac{\omega_1(\vec q)+\omega_2(\vec q)}
{E^2-(\omega_1(\vec q)+\omega_2(\vec q))^2+i\epsilon},
\non 
\omega_{1,2}(\vec q)&=&\sqrt{m_{1,2}^2+\vec q^2}\, ,
\label{prop_cont}
\ea
with meson masses $m_1,\,m_2$ in channel $j$ (in our particular case, $I=0$,
the masses  $m_1$ and $m_2$ are equal. To ease notations,  below we suppress
the index $j$ in the masses, etc.). We call $K\bar K$ channel 1 and $\pi\pi$
channel 2,  and $E=\sqrt{s}$ is the center-of-mass (CM) energy. The function
$G$  requires regularization and we use, as in \cite{npa}, a cutoff 
regularization with a natural-size cutoff  $q_{\rm max}=904$ MeV~\cite{npa},
though it is also  customary to use dimensional regularization with a
subtraction constant to  regularize the loops \cite{ollerulf}. The choice of a
particular  regularization scheme does not, of course, affect our
argumentation.  Eq.~(\ref{prop_cont}) is just the relativistic generalization,
consistent  also with the meson statistics, of the integral of the ordinary 
non-relativistic Green's function $1/(E-H_0)$ in the Lippmann Schwinger 
equation. By using the above cutoff,  one obtains very good results for the
$\pi \pi$ and $K \bar{K}$  scattering amplitudes up to about $\sqrt{s}\sim
1200$ MeV \cite{npa}.

We put now the same model in a finite cubic box of side length $L$  and predict
the discrete spectrum that emerges in such a box.  Neglecting all partial waves
except the S-wave~\footnote{As rotational symmetry is broken in the box,
partial waves mix in general, e.g. $S$-wave mixes with the $L=4$ partial wave
(G-wave). This effect is expected to be small and neglected in this study.},
the only change to be made is to substitute  the function $G$ in Eq.
(\ref{bse})  by  $\tilde G={\rm diag}\,(\tilde G_1,\tilde G_2)$, where 
\ba
\tilde G_{j}&=&\frac{1}{L^3}\sum_{\vec q}^{|\vec q|<q_{\rm max}}
\frac{1}{2\omega_1(\vec q)\,\omega_2(\vec q)}\,\,
\frac{\omega_1(\vec q)+\omega_2(\vec q)}
{E^2-(\omega_1(\vec q)+\omega_2(\vec q))^2},
\non 
\vec q&=&\frac{2\pi}{L}\,\vec n,
\quad\vec n\in \mathds{Z}^3 \ ,
\label{tildeg}
\ea
and again, $\omega_{1,2}(\vec q)=\sqrt{m_{1,2}^2+\vec q^2}$. In other words,
the only change is to replace the integral over the continuous variable $\vec
q$ in Eq. (\ref{prop_cont}) by a sum over the discrete values, corresponding
to  periodic b.c.\ . 

The discrete spectrum of the system is given by the poles of the scattering
$T$-matrix $\tilde T(E)$  in a finite volume.  It is interesting to show the
relation of the present approach to the one of L\"uscher for the case of one
channel. To this end, assume first that the energy $E$ is above threshold, and
take into account the fact that the potential $V$ is the same in a finite and
in  the infinite volume, up to  exponentially suppressed terms. The poles
emerge at the energies $\tilde T^{-1}(E)=0$, i.e.,
\be
V^{-1}(E)-\tilde G(E)=0\, .
\label{vmin1}
\ee 
It can be easily seen that the above equation produces an infinite tower  of
discrete levels.

Further, using Eq.~(\ref{vmin1}), the scattering matrix in the infinite volume
can be rewritten, for the discrete eigenenergies satisfying Eq.~(\ref{vmin1}),
as
\be
T(E)=\left(V^{-1}(E)-G(E)\right)^{-1}= \left(\tilde G(E)-G(E)\right)^{-1} \ . 
\label{extracted_1_channel}
\ee
Below, we shall use the normalization of Ref.~\cite{npa}. In this
normalization, the infinite-volume scattering matrix $T(E)$ is related to the
phase shift by
\be
T(E)=\frac{-8\pi E}{p\cot \delta(p)-i\,p}\, ,
\ee
where  $p=\lambda^{1/2}(E^2,m_1^2,m_2^2)/(2E)$ and $\lambda(x,y,z)$ stands for 
the K\"all\'en triangle function. Substituting this into
Eq.~(\ref{extracted_1_channel}), we arrive at
\ba
\label{eq:pre_luescher}
p\cot\delta(p)=-8\pi E\biggl\{\tilde G(E)-\biggl(G(E)+\frac{ip}{8\pi E}\biggr)
\biggr\}
\non
\mbox{[ above threshold ]}\, .
\ea
Here, we suppressed the index ``$j$'' in the Green function, because the
one-channel problem is considered. Note also that, for $E$ above threshold,  we
get $G(E)+ip/(8\pi E)={\rm Re}\,G(E)$, and the r.h.s. of
Eq.~(\ref{eq:pre_luescher}) is real, as it should.

Finally, we perform the analytic continuation of Eq.~(\ref{eq:pre_luescher})
below threshold. The l.h.s. of this equation is defined only above threshold.
However, it is known that the l.h.s. obeys the effective-range expansion near
threshold
\be
\label{eq:effrange}
p\cot\delta(p)=-\frac{1}{a}+\frac{1}{2}\,rp^2+O(p^4)\, ,
\ee 
where $a$ and $r$ denote the scattering length and the effective range, 
respectively. Note also  that this is an expansion in the variable $p^2$ and
hence the threshold  is a regular point in this expansion. Performing now the
analytic  continuation  below threshold, where $p=i\gamma$, we get
\begin{multline}
\label{eq:pre_luescher1}
 -\frac{1}{a}-\frac{1}{2}\,r\gamma^2+\cdots=-8\pi E\biggl\{\tilde G(E)
-\biggl(G(E)-\frac{\gamma}{8\pi E}\biggr)\biggr\}
\\
\mbox{[ below threshold ]}
\, .
\end{multline}  	

\subsection{Relation to the L\"uscher equation}
\label{sec:relation}

Next, we wish to demonstrate that Eqs.~(\ref{eq:pre_luescher}) and
(\ref{eq:pre_luescher1}) are nothing but the ordinary L\"uscher equation in
regions of energy and $L$  where the L\"uscher approach is a good 
approximation (see also Ref.~\cite{luscher}). To this end, let us use the
identity 
\begin{multline}
\frac{1}{2\,\omega_1\,\omega_2}\,\frac{\omega_1+\omega_2}
{E^2-(\omega_1+\omega_2)^2+i\epsilon}\\
=\frac{1}{2E}\,\frac{1}{p^2-\vec q^2+i\epsilon}
-\frac{1}{2\omega_1\,\omega_2}\,\frac{1}{\omega_1+\omega_2+E}\\
-\frac{1}{4\,\omega_1\,\omega_2}\,\frac{1}{\omega_1-\omega_2-E}
-\frac{1}{4\,\omega_1\,\omega_2}\,\frac{1}{\omega_2-\omega_1-E}
\label{lalala}
\end{multline}
(below threshold, the $i\epsilon$ prescription should be omitted). This
identity should be used both in $\tilde G(E)$ and $G(E)$.

Let us now take into account the fact that the singularities of the last three
terms in the r.h.s. of Eq.~(\ref{lalala}) are not located in the scattering
region ($E$ above threshold) and are separated from the threshold by a distance
of the order of the particle masses $m_{1,2}$. According to the regular
summation theorem~\cite{luscher,Luscher:1990ux} for such functions, the
difference between the sum and the integral is  exponentially suppressed at
large volumes. Consequently, the contribution from the last three terms is
exponentially suppressed  in the difference  $\tilde G(E)-G(E)$ and can be
neglected. This difference above threshold takes the form
\begin{multline}
\label{eq:equiv}
\tilde G(E)-G(E)=\biggl\{\frac{1}{L^3}\sum_{\vec q}^{|\vec q|<q_{\rm max}}
-\int^{|\vec q|<q_{\rm max}}\frac{d^3\vec q}{(2\pi)^3}\,\biggr\}
\\
\times\,\frac{1}{2E}\frac{1}{p^2-\vec q^2+i\epsilon}
+\cdots
=\frac{1}{2E}\,\frac{1}{L^3}\sum_{\vec q}^{|\vec q|<q_{\rm max}}
\frac{1}{p^2-\vec q^2}
\\
+\frac{1}{4\pi^2E}\,
\left(q_{\rm max}+\frac{p}{2}\log\frac{q_{\rm max}-p}{q_{\rm max}+p}\right)
+\frac{ip}{8\pi\,E}+\cdots\, ,
\end{multline}
where the ellipses stand for the exponentially suppressed terms.  Below
threshold, the last term in the r.h.s. gets replaced by  $-\gamma/(8\pi E)$.

Moreover, as seen from Eq.~(\ref{eq:equiv}), one may in fact remove here the
cutoff, sending $q_{\rm max}\to\infty$.  Indeed, one should obviously take a
$q_{\rm max}$ such that $p^2< q_{\rm max}^2$ in the whole region of interest to
us. If we sum and integrate from $q_{\rm max}$ to $q_{\rm max}'$, with $q_{\rm
max}'>q_{\rm max}$, the denominator $p^2-\vec q^2$ is  not singular and,
according to the regular summation theorem, only exponentially suppressed
corrections may arise. Finally, noting that (see, e.g., Ref.~\cite{beane})
\begin{multline}
\label{eq:Z00}
\lim_{q_{\rm max}\to\infty}\biggl\{\frac{1}{L^3}
\sum_{\vec q}^{|\vec q|<q_{\rm max}}
\frac{1}{p^2-\vec q^2}-\frac{q_{\rm max}}{2\pi^2}\biggr\}
\\
=-\frac{1}{2\pi^{3/2}L}\,{\cal Z}_{00}(1,\hat p^2)\, ,\quad\quad
\hat p=\frac{pL}{2\pi}\, ,
\end{multline}
where ${\cal Z}_{00}$ stands for the L\"uscher zeta-function, we can directly
verify that Eq.~(\ref{eq:pre_luescher}) (or its extrapolation below threshold,
Eq. (\ref{eq:pre_luescher1})) indeed coincides with the L\"uscher equation 
\be
p\cot\delta(p)=\frac{2\pi}{L}\,\pi^{-3/2}\,{\cal Z}_{00}(1,\hat p^2)
\ee
and is cutoff-independent up to exponentially small  corrections\footnote{Note
that, despite the non-relativistic appearance of the denominator in
Eqs.~(\ref{eq:equiv}) and (\ref{eq:Z00}) that at the first glance looks like a
finite-volume version of UCHPT with non-relativistic
propagators~\cite{junkowave,Gamermann:2009uq}, the formalism of course
possesses the relativistic dispersion law.}.

Equation~(\ref{eq:Z00}) can be used for practical purposes to evaluate the
L\"uscher function, if one wishes, and as noticed  in Ref.~\cite{beane} its
finiteness is explicitly shown.  However, for practical purposes it is more
convenient  to keep the $\log$ terms in Eq.~(\ref{eq:equiv}),  which guarantees
a faster convergence, and we obtain  ($\Lambda=q_{\rm max}L/(2\pi)$)
\ba
&&{\cal Z}_{00}(1,\hat p^2)=\lim_{\Lambda\to\infty}\left(\frac{1}{\sqrt{4\pi}} 
\sum_{\vec n}^\Lambda \frac{1}{\vec n^2-\hat p^{\,2}} 
- F(\hat p^{\,2},\Lambda)\right) \ ,
\non &&
F(\hat p^{\,2},\Lambda)=\sqrt{4\pi}\,\Lambda
\non &\times&
\begin{cases}
\left(1+\frac{\hat p_\Lambda}{2}\log\frac{1-\hat p_\Lambda}
{1+\hat p_\Lambda}\right),
&\hat p_\Lambda=\hat p/\Lambda \text{ for }\hat p^{\,2}>0\\
\left(1+\pi\kappa/2-\kappa\,\text{atan}\, 1/\kappa\right), 
&\kappa=\sqrt{-\hat p^{\,2}}/ \Lambda\text{ else. }
\end{cases}
\non
\label{easyzoo}
\ea
Below threshold, we have subtracted from $G$ the analytic extrapolation of 
$i\,{\rm Im}\,G$. The convergence in practice is found for values of the
dimensionless $\Lambda$ as low as $\Lambda\sim \hat p+3$. The convergence is
mostly limited by the  artefacts introduced by the sharp cut-off $\Lambda$. It
can be improved by averaging over several values of $\Lambda>\hat p+3$. If
one omits the log term in Eq.~(\ref{easyzoo}) like in Ref.~\cite{beane}, the
same level of convergence is only reached for larger values of $\Lambda>\hat
p + 6$.

We further comment on the question,  whether the finite-volume corrections,
stemming from the last three terms in Eq.~(\ref{lalala}), are uniformly 
suppressed by an exponential factor in the whole energy region  measured on the
lattice. From Eq.~(\ref{lalala}) it is seen that, e.g., in the equal-mass case,
the third and the fourth terms on the r.h.s. have a pole at $E=0$, so the
convergence is not uniform in the vicinity of this point  (note that the
left-hand side of Eq.~(\ref{lalala}) shows no singularity at $E=0$). The
singularity of the second term of the right-hand side is located further left in
the energy plane. In general, for the case of non-equal masses, the
singularities arise for the energies $E<|m_1-m_2|$. Of course, in case of
one-channel scattering,  this point lies below threshold, in the unphysical
region.  If the energy is taken larger, one may safely use the  regular
summation theorem and show that the contribution of the last three terms to the
difference $G-\tilde G$ is uniformly suppressed by an exponential factor. For
smaller energies, this is no more true. Yet, for a case like $\pi\Sigma$, the
singularity at $|m_1-m_2|$ is situated $2M_\pi$ below threshold and corrections
due to the neglected terms show up much before the pole. One should also note
that at energies above threshold, even if these neglected terms are
exponentially suppressed, one still finds a contribution from those terms and
their numerical relevance must be checked. This is done in
appendix~\ref{sec:appendix}.

In addition to this, one can readily see  that there is a potential problem if
one considers the multi-channel L\"uscher equation. Let $m_1,m_2$ and
$m_1',m_2'$ be the masses in the  ``heavy'' and ``light'' channels,
respectively, so that $m_1+m_2>m_1'+m_2'$.  If $|m_1-m_2|>m_1'+m_2'$, the
singularity from the ``heavy'' channel comes into the physical region of the
light channel and using the decomposition (10) can not be justified any more.
Fortunately, it does not happen in the case of $\pi\pi-K\bar K$ and
$\pi\eta-K\bar K$  coupled-channel scattering, which is considered in the
present paper.

Even if in the particular case of interest the redundant poles, which stem  from
the decomposition given in Eq. (10), do not sneak into the physical  region, it
is still legitimate to ask, whether the presence of such nearby singularities
may affect the numerical accuracy of the method (even above threshold, as we
pointed out above).  We would like to stress the following once more: the
difference which we are trying to estimate, is exponentially suppressed at large
$L$ and thus beyond the accuracy claimed by the L\"uscher  approach. However, as
we can see in Fig. 1 below, the volumes for which the energies of the low-lying
levels are of order of 1~GeV or less, are indeed not that large. 

\subsection{Cutoff effects}

Finally, we wish to comment on the artifacts coming from the finite cutoff
$q_{\rm max}$. Keeping the sharp cutoff of natural size yields small unphysical
discontinuities in the predicted energy levels. These discontinuities disappear
if we use a smooth cutoff instead, 
\ba
\tilde G_j&\to&\tilde G_{S,j}+(G_j-G_{S,j})\non
G_{S,j}&=&\int\frac{d^3\vec q}{(2\pi)^3}\frac{F(|\vec q|)}
{2\omega_1(\vec q)\,\omega_2(\vec q)}\,\,
\frac{\omega_1(\vec q)+\omega_2(\vec q)}
{E^2-(\omega_1(\vec q)+\omega_2(\vec q))^2+i\epsilon}\, ,\non
\tilde G_{S,j}&=&\frac{1}{L^3}\sum_{\vec q}
\frac{F(|\vec q|)}{2\omega_1(\vec q)\,\omega_2(\vec q)}\,\,
\frac{\omega_1(\vec q)+\omega_2(\vec q)}
{E^2-(\omega_1(\vec q)+\omega_2(\vec q))^2}\, .
\label{tildegs}
\ea
Here, 
\ba
F(q)=\frac{a^{m}}{a^{m} + q^{m}}\, .
\label{fofa}
\ea
We choose $a=1.2$ GeV, $m=12$. This form factor is almost equal to one up
to approximately $q\simeq 800$ MeV and then drops smoothly to zero  at
around $q\simeq 1.7$~GeV.

The only caveat in this procedure is that the form factor $F$ must be equal to
one for the on-shell value of the momentum $q$, in order to obey   unitarity
exactly. In choosing $F$, we have demanded that it is different from one by
less than  $1\permil$ for all energies considered here.

\section{Two-channel formalism}
\label{sec:twisted-asym}

\subsection{General setting}

\begin{figure}
\includegraphics[width=0.38\textwidth]{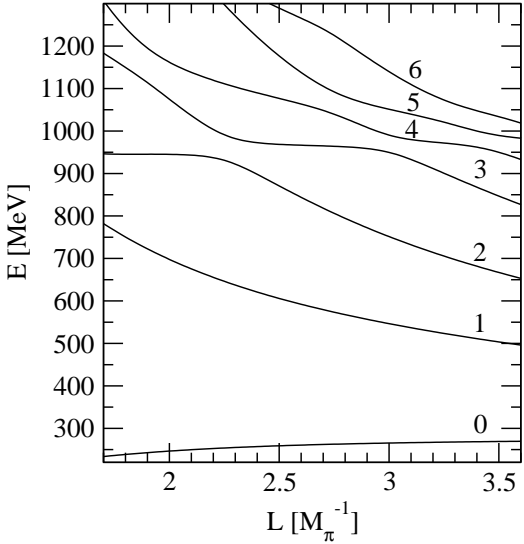}
\caption{Energy levels as functions of the cubic box size $L$, derived 
from the chiral unitary approach of Ref.~\cite{npa} and using 
$\,\tilde G$ from Eq.~(\ref{tildegs}).}
\label{fig:levels_nonsmooth}
\end{figure}

The generalization of the L\"uscher formalism to  coupled-channel scattering is
straightforward. The secular equation that determines the spectrum is given by
\be
\label{eq:det}
\det(\mathds{1}-V\tilde G)=1-V_{11}\tilde G_1-V_{22}\tilde G_2
+(V_{11}V_{22}-V_{12}^2)\tilde G_1\tilde G_2=0\, .
\ee
The energy levels which have been determined from this equation, using the
input potentials $V_{ij}$ from Ref.~\cite{npa}, are shown in 
Fig.~\ref{fig:levels_nonsmooth}.

In Refs.~\cite{Lage:2009zv,akaki}, we have introduced a very useful quantity
called {\it pseudophase}. We shall often refer to this quantity below, because
it is convenient in the discussion of the L\"uscher equation. By definition, the
pseudophase is the phase extracted from a given energy level, using the {\it
one-channel} L\"uscher equation~(\ref{eq:pre_luescher}) (i.e., neglecting the
channel with a higher threshold). From the definition, it is clear that, well
below the inelastic threshold, the pseudophase coincides with the usually
defined scattering phase, up to the exponentially suppressed finite-volume
corrections. Above the inelastic threshold, a tower of resonances emer\-ges in
the pseudophase, which reflect the opening of the second channel.

\begin{figure}
\includegraphics[width=0.38\textwidth]{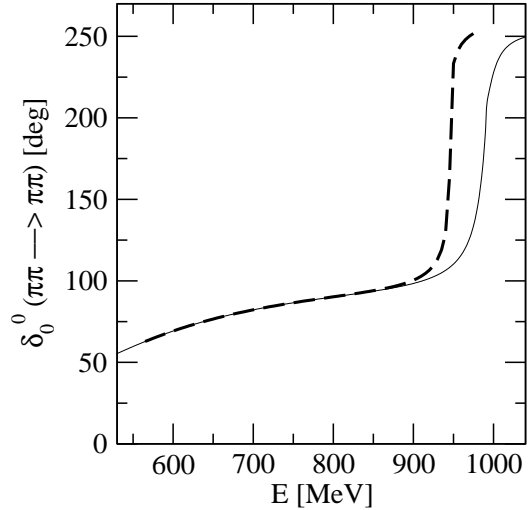}
\caption{Solid line: The original $\pi\pi$ scattering S-wave phase shift 
$\delta_0^0$ from the 
chiral unitary amplitude of Ref.~\cite{npa}; Dashed line: 
the pseudophase extracted from the spectrum of 
Fig.~\ref{fig:levels_nonsmooth}, using the one-channel 
equation~(\ref{eq:pre_luescher}) for the 
level 2.}
\label{fig:extracted_one_channel}
\end{figure}

In Fig.~\ref{fig:extracted_one_channel} we show the $\pi\pi$ pseudophase, 
extracted from the energy levels in Fig.~\ref{fig:levels_nonsmooth}. It is seen
that, below 900 MeV, the pseudophase is very close to the $\pi\pi$ scattering
phase $\delta_0^0$. Above 900~MeV, the pseudophase rises faster than 
$\delta_0^0$, owing to the presence of the $K\bar K$ threshold.

\subsection{Multi-channel approach}
\label{subsec:multi}

Next, we turn to the central problem for the two-channel (generally,  the
multi-channel) approach. As mentioned already, unlike the one-channel case, a
single measurement of  a value of the energy $E(L)$ at a given $L$ can not
provide the full information about three independent quantities $V_{ij}(E(L))$
at the same energy. The  variation of $L$ does not help, because the energy
also changes. In order to determine all $V_{ij}$ independently (and, thus, the
pole positions, which are eventually determined by the $V_{ij}$), there are
several options:

\begin{itemize}
\item[i)] 
In Ref.~\cite{akaki} we have proposed to use twisted b.c. for the $s$-quark. If
$\vec \theta$ with $0<\theta_i<2\pi,~i=1,2,3$ denotes the twisting angle,
imposing the twisted b.c. is equivalent to adding $\pm\vec\theta/L$ to the
relative three-momentum of a kaon (anti-kaon)  in the CM frame, whereas the
relative momenta in the $\pi\pi$ intermediate state do not change (Below, for
simplicity,  we shall always use the symmetric choice $\theta_i=\theta$).
Consequently, the expression of the $\pi\pi$ loop function does not change,
whereas the $K\bar K$ loop function becomes $\theta$-dependent (the pertinent
expressions can be found in Ref.~\cite{akaki}). Such a procedure is very
convenient, because it allows to move the $K\bar K$  threshold and thus perform
a detailed scan of the region around 1~GeV, where the $f_0(980)$ is located.

The idea behind using twisted b.c. is that in this case one has two
``external'' parameters $L$ and $\theta$, which can be  varied without changing
the dynamics of the system. In particular, one may adjust these so that a given
energy level does not move. On the other hand, since the loop functions,
entering the secular equation (\ref{eq:det}), depend on $L$ and $\theta$, for
different values of these parameters one gets a (non-degenerate) system of
equations that allow the determination of the individual matrix elements
$V_{ij}$ at a given energy.

For illustrative purposes, in Fig.~\ref{fig:levels_smoothened and with twisted}
we display the $\theta$-dependent spectrum for $\theta=0$ (periodic b.c.),
$\theta=\pi/2$ and $\theta=\pi$  (antiperiodic b.c.), obtained by using
UCHPT in a finite volume. It is seen that the spectrum is strongly sensitive to
twisting  at the energies around 1~GeV, which is exactly the energy region  we
are interested in.

We realize that it could be quite challenging to implement this idea (with a
complete twisting, including the sea quarks) in present-day lattice
simulations. A partial twisting (of only the valence quarks) can be performed
more easily. The question, whether such a partial twisting will still enable
one to extract useful information about the properties of the multi-channel
resonances, is still open. We plan to address this (very interesting) question
within the EFT framework in the future.

\begin{figure}
\includegraphics[width=0.38\textwidth]{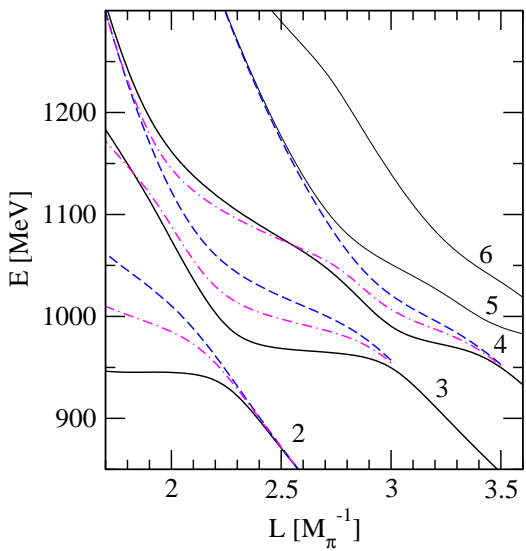}
\caption{
Solid lines: Spectrum $E(L)$ as in Fig.~\ref{fig:levels_nonsmooth};
 Dashed lines: Applying twisted b.c. $\theta=\pi$ as in Ref.~\cite{akaki}, 
only for the $K\bar K$ channel and for the levels 2,3, and 4;
 Dashed dotted lines: with $\theta=\pi/2$.}
\label{fig:levels_smoothened and with twisted}
\end{figure}

\item[ii)]
Introducing the twisting angle $\theta$ gives us an additional adjustable
parameter that allows one to determine the individual matrix elements $V_{ij}$.
The same goal can be achieved without twisting, performing the calculations on
the asymmetric lattices $L_x\times L_y\times L_z$ where, generally, $L_{x,y,z}$
are not all equal~\footnote{Effects of additional partial wave mixing are
neglected here.}. Note that scattering in an asymmetric box has  already been
studied on the lattice, see Refs.~\cite{Li:2007ey,Liu:2007qq,Liu2,Ishii}.

Suppose $L_{x,y,z}$ have been adjusted so that the value of the energy
$E(L_{x,y,z})=E$ stays put. In practice one evaluates the trajectories of the
levels, and cuts them by the line of $E=\text{const}$ to determine the $L_i$.
The loop functions $\tilde G_j$ depend both on the energy $E$ and the box
configuration given by $L_{x,y,z}$. Note that $L^3$ in Eq.~(\ref{tildegs}) is
now replaced by $L_xL_yL_z$ and $q_i=2\pi n_i/L_i$, $i=1,2,3$, $n_i\in
\mathds{Z}$. We label the different configurations with the same energy by an
index $a=1,2,\cdots$, and denote  the pertinent values of the loop functions by
$\tilde G^{(a)}(E)$. Performing the measurement for three different
configurations and solving the secular equation, one can determine the matrix
elements of the potential separately,
\ba
\left(
\begin{array}{l}
V_{11}\\
V_{22}\\
V_{12}^2-V_{11}V_{22}
\end{array}
\right)
=\tilde G_L^{-1}
\left(
\begin{array}{l}
1\\
1\\
1
\end{array}
\right),
\label{twochann}
\ea
where
\ba
\tilde G_L=
\left(
\begin{array}{lll}
\tilde G_1^{(1)}\hspace*{0.3cm}&\tilde G_2^{(1)}\hspace*{0.3cm}&
\tilde G_1^{(1)}\tilde G_2^{(1)}\\
\tilde G_1^{(2)}&\tilde G_2^{(2)}&\tilde G_1^{(2)}\tilde G_2^{(2)}\\
\tilde G_1^{(3)}&\tilde G_2^{(3)}&\tilde G_1^{(3)}\tilde G_2^{(3)}
\end{array}
\right).
\ea
In Fig.~\ref{fig:levels_asymmetric} we show a particular energy level in the
vicinity of 1 GeV,  calculated for different configurations of the asymmetric
box. It is seen that the level is not very sensitive to the configuration. For
this reason, a larger error bar is {\it a priori} expected, if observables  are
extracted from these data.

\begin{figure}
\includegraphics[width=0.38\textwidth]{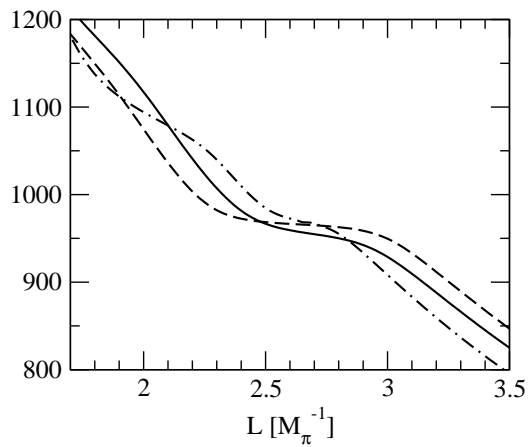}
\caption{Energy level 3 in an asymmetric box of dimensions $(L_x=L,
 L_y=L, L_z=x\,L)$, where $x=0.6$ (solid lines), $x=1.0$ (dashed lines),
 $x=1.4$ (dash-dotted lines).}
\label{fig:levels_asymmetric}
\end{figure}

For illustrative purposes, in Fig.~\ref{fig:extracted_2_channels_2} we display
different physical observables in the $\pi\pi$ and $K\bar K$ scattering, that
can be extracted from the energy levels by using the above-mentioned methods,
no errors yet assigned. As seen, both methods have in principle the capability
to address the problem. 

\begin{figure*}
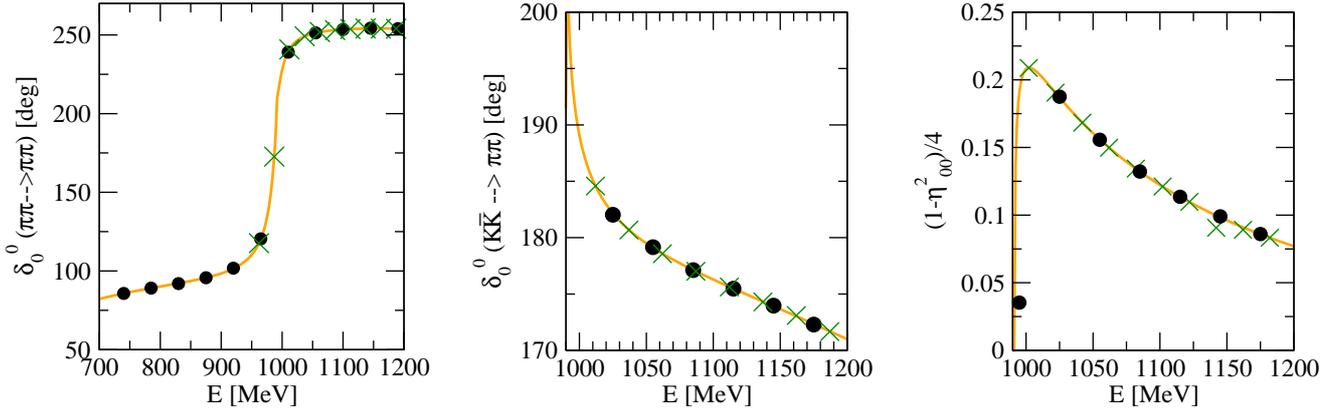

\includegraphics[height=0.3\textwidth]{extracted_2_channels_newscheme.eps} 
\hspace*{0.3cm}
\includegraphics[height=0.3\textwidth]
{extracted_2_channels_newscheme_del_1+2.eps}
\hspace*{0.3cm}
\includegraphics[height=0.3\textwidth]{eta_newscheme.eps}
\caption{Left: $\pi\pi\to\pi\pi$ phase shift. Center and right: $K\bar
K\to\pi\pi$ phase shift and inelasticity. Solid line: infinite-volume phase
shift from which the lattice spectrum $E(L)$ is calculated; Crosses (X):
Reconstructed phase shift using the twisted b.c. with $\theta=0,\,\pi/2,\,\pi$
of Fig.~\ref{fig:levels_smoothened and with twisted};  Solid circles: Using the
3 asymmetric lattice levels of Fig.~\ref{fig:levels_asymmetric}.}
\label{fig:extracted_2_channels_2}
\end{figure*}

\end{itemize}

For completeness, one should mention that measuring at least 3 different
excited levels {\it at the same energy but different values of $L$} could, in
principle, also achieve the goal. Such a proposal, however, looks quite
unrealistic in the view of the present state of art in the scalar meson
sector. 

\subsection{Threshold effects}

At this place, we would like to comment on the threshold effects, which have a 
capability to seriously complicate the search of the near-threshold resonances
-- exactly the task we are after. The resonances on the lattice reveal
themselves in the form of a peculiar behavior of the volume-dependent energy
levels near the resonance energy called ``avoided level
crossing''~\cite{Wiese:1988qy}. For example, in 
Fig.~\ref{fig:levels_smoothened and with twisted}, the solid curves (levels,
corresponding to the periodic b.c.), display a perfect avoided level crossing
slightly below 1 GeV. Does this signal the presence of the $f_0(980)$?

In order to answer this question, it suffices to change the parameters of the
potential slightly, making the $f_0(980)$ disappear. This can be achieved, for
example, by merely reducing the strength of the potential $V_{11}$ that
describes the scattering in the  $K\bar K\to K\bar K$ channel. Namely, replacing
$V_{11}\to \eta V_{11}$ and varying $\eta$ between 1 and 0, it is seen that the
$f_0(980)$ disappears already at $\eta\simeq 0.8$ (the pole moves to a hidden
sheet). Below, we shall study the dependence of the energy levels on the
parameter $\eta$. 

\begin{figure}
\includegraphics[width=.48\textwidth]{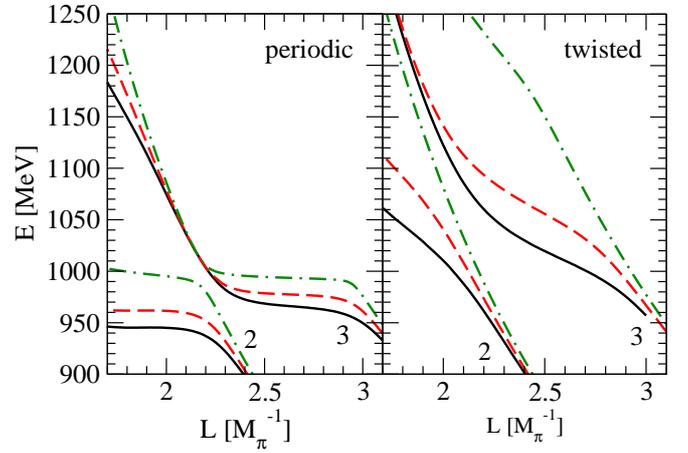}
\caption{Energy levels, different values of $\eta$.
Left panel corresponds to periodic b.c.
$\theta=0$, right panel to twisted b.c. $\theta=\pi$.
Solid lines: $\eta=1$; Dashed lines: $\eta=0.8$, Dash-dotted lines: $\eta=0$. 
}  

\label{fig:deception-energy}
\end{figure}

\begin{figure}
\includegraphics[width=.48\textwidth]{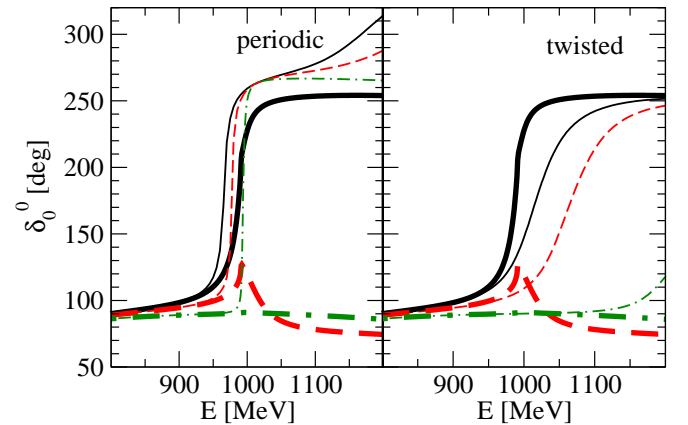}
\caption{Thick lines: true $\pi\pi$ scattering phases, different values of
$\eta$; Thin lines: pseudophases, different values of $\eta$. Left panel
corresponds to periodic b.c. $\theta=0$, right panel to twisted b.c.
$\theta=\pi$. Solid lines: $\eta=1$; Dashed lines: $\eta=0.8$, Dash-dotted
lines: $\eta=0$. }  
\label{fig:deception-phase}
\end{figure}

Let us first stick to the periodic b.c.\ . In the left panel of 
Fig.~\ref{fig:deception-energy} one immediately sees that the avoided level
crossing in the energy levels persists even for vanishing $\eta$, when there
is  no trace of the $f_0(980)$ any more in the scattering phase (see
Fig.~\ref{fig:deception-phase}). This shows very clearly that the avoided level
crossing in this case has nothing to do with the  $f_0(980)$ and is due to the
presence of the $K\bar K$ threshold. The modified problem has memory of the
$K\bar K$ channel because the $V_{12}$ term is still kept and connects the two
channels.

The pseudophase which is displayed in the left panel of
Fig.~\ref{fig:deception-phase}  also demonstrates this statement. We see that
it barely changes with $\eta\to 0$, whereas the true scattering phase is
subject to dramatic changes. A rapid change of the pseudophase by $\pi$ around
1~GeV is related to the presence of the $K\bar K$ threshold and not to  the
$f_0(980)$.

Now, the essence of the problem becomes clear. It seems that the presence of
the threshold masks the presence of the resonance, and the energy levels become
largely insensitive to the latter. How accurate the lattice data should be in
order to provide a reliable extraction of the resonance parameters in the
vicinity of a threshold?

It should be pointed out that the use of twisted b.c. gives one possible
solution to the above problem. Namely, as already  mentioned above, as the
twisting angle $\theta$ increases, the $K\bar K$ threshold moves up, whereas
the $f_0(980)$ stays put. Consequently, for larger values of $\theta$, these
two are well separated, and the pseudophase reveals a larger dependence on the
parameter $\eta$. This is demonstrated in the right panels of
Figs.~\ref{fig:deception-energy} and  \ref{fig:deception-phase}. 

It remains to be seen, whether the use of twisted b.c. could be advantageous
for resolving near-threshold resonances. Below, we address this issue by means
of fits to  synthetic lattice data.

\subsection{The framework based on UCHPT}

Finally, we would like to turn to the framework of UCHPT in a finite
volume, which is a central issue considered in the present paper. In
section~\ref{subsec:multi}, we did not assume any specific ansatz for the
potential $V_{ij}$ which  is being extracted from the data. It is clear,
however, that, changing the box geometry and/or the twisting angle, we can not
arrive exactly to the same energy, and some kind of an interpolation procedure
will be  needed. This could be achieved by evaluating many points belonging to
the same level for different values of $L$ and performing a best fit to the
data which will return an accurate trajectory with smaller errors than the
individual ones (see, e.g., Refs.~\cite{Bayes,Bayes2,Chen})~\footnote{We are aware
that present day lattice simulations do not provide more than a few volumes.}.

Alternatively, one may choose the parameterization of the potential $V_{ij}(E)$
as in UCHPT (e.g., as in Eq.~(\ref{fitv}) below),  and then fitting the
parameters of $V_{ij}$ to the data on the lattice introduces certain
model-dependence, but provides a fair input based on the successful UCHPT that
allows the interpolation of the lattice data. With more lattice data available,
$V_{ij}$  as functions of the energy can be fitted more accurately, making the 
initial ansatz for it less relevant. We would like to stress that the present
ansatz for the potential, in our opinion, is simple, rather general and
theoretically solid. As mentioned above, it has been successfully used in the
past to accurately describe the scattering in the $\pi\pi/\pi\eta$ and $K\bar K$
systems (at physical values of the quark masses), see 
Refs.~\cite{npa,Pelaez:2003dy,Kaiser:1998fi,Locher:1997gr}. This means that the
effect of the higher-order terms in the potential should be small. Of course,
this assumption could be checked by incorporating such higher order terms as it
has been done e.g. in the case of $K^-p$ scattering \cite{Borasoy:2006sr}.  It
can be expected that, in the first approximation,  the functional form of the
potential stays intact when we go to higher quark masses that are used in 
present lattice simulations (the numerical values of the parameters may depend
on  the quark masses). 

Below, performing the fits to the synthetic lattice data with errors, we shall
explicitly demonstrate how the approach works.

\section{Fit to the energy levels}
\label{sec:fit}

\subsection{Twisted boundary conditions versus asymmetric boxes}
\label{sec:error1}

\begin{figure}
\includegraphics[width=0.38\textwidth]{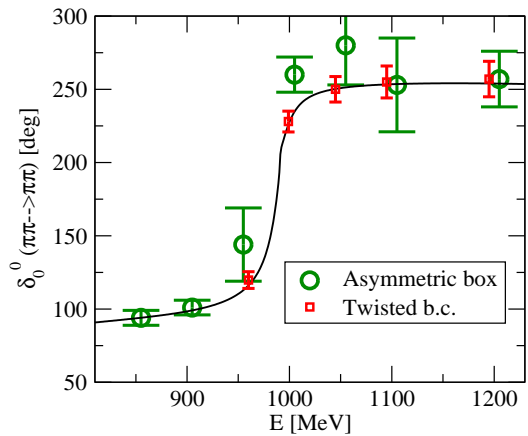}
\caption{Error estimates for two different extraction schemes.} 
\label{fig:errors}
\end{figure}

Below, we shall carry out a preliminary numerical estimate of the efficiency of
the two different schemes which were discussed above. To this end, first we
produce the energy levels by using UCHPT in a finite volume.  Then we assume
that each data point $E(L)$ is produced with some uncertainty $\Delta E(L)$.
Reversing the dependence on $L$, one obtains $L(E)$ and $\Delta L(E)$ (in case
of an asymmetric box we have taken $L_i=\alpha_i\,L$ for three different
choices of $\alpha_i$ and we take the same $\Delta L$ as in the symmetric
case). In general, for a given $E(L)$ and $\Delta E(L)$, the inverse $\Delta
L(E)$ is a very complicated function of energy, depending on the tangent of the
curve. Since we are making only a rough estimate here, we simplify our task by
assuming a {\it constant} error  $\Delta L=0.02 M_\pi^{-1}$. Further, using the
von Neumann rejection method, we generate a Gaussian distribution for the
variable $L$ centered around the exact solution $a=L(E)$
\ba
f(L)=\frac{1}{\Delta L\sqrt{2\pi}}\exp\left\{
-\frac{(L-a)^2}{2(\Delta L)^2}\right\}\, .
\label{gauss}
\ea
We perform the calculations of the infinite-volume phase shifts for each value
of $L$ and determine the average value and uncertainty, using
Eq.~(\ref{twochann}).  The results for two different methods are shown in
Fig.~\ref{fig:errors}.
Below and above the $K\bar K$ threshold, both methods deliver  very similar
phase shifts with a comparable error. Around the $K\bar K$ threshold,  as
expected, the use of twisted b.c. produces phase shifts  with an error smaller
than in the other method.

\subsection{Extraction of the $f_0(980)$ pole position from the data}

After the preliminary study of the different schemes carried out in the
previous subsection, we turn to the main topic of our paper and perform the fit
of the data by using UCHPT in a finite volume. To this end, we consider the
procedure for producing various synthetic data sets and the strategy of the
fit.

\begin{figure}
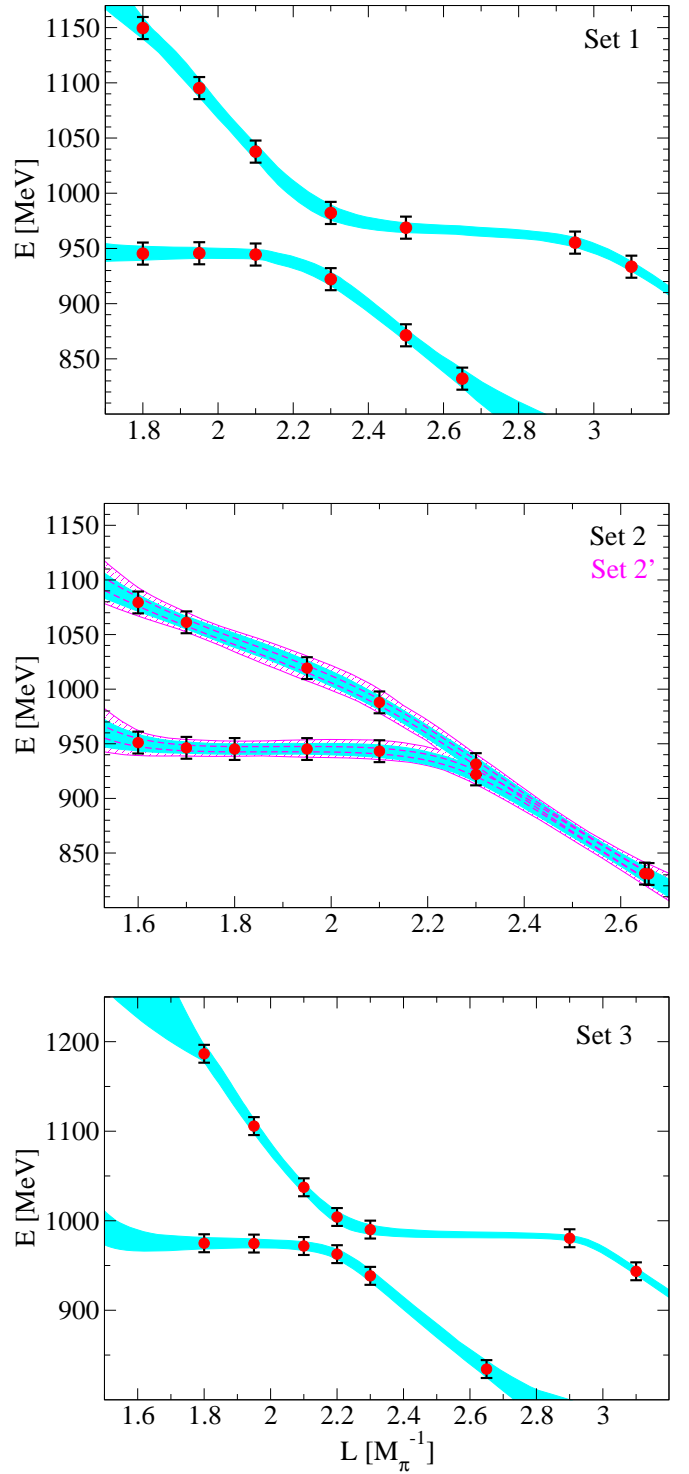

\includegraphics[width=0.48\textwidth]{levels_w_fo_without_twisted_shapes.eps}\\

\vspace*{0.4cm}

\includegraphics[width=0.48\textwidth]{levels_w_fo_mixed_BC.eps}\\

\vspace*{0.4cm}

\includegraphics[width=0.48\textwidth]{levels_without_fo_without_twisted_shape.eps}\\
\caption{Generated data points (10 MeV error, 13 points for each set).  Upper
figure: using levels 2 and 3 with periodic b.c.;  Central figure: using level 2
with periodic and antiperiodic b.c.; Lower figure: case without $f_0(980)$ 
(using levels 2,3 with periodic b.c.).  Fits that fulfill the $\chi_{\rm
best}^2+1$ criterion are also shown in all figures [solid (cyan) bands].  The
set $2'$ corresponds to data with displaced centroids (see
Sec.~\ref{sec:twoprime} for more details). In the central figure, the fits to
set $2'$ are shown by dashed lines and the hatched (magenta) bands.}
\label{fig:errors2}
\end{figure}

The central values $E_n(L),~n=0,1,\cdots$ are produced by solving the secular
equation (\ref{eq:det}) with the potentials from UCHPT, by using 
periodic/antiperiodic b.c. (below, we do not study the case of the asymmetric
boxes). At the next step, a constant error  $\Delta E=10~\mbox{MeV}$ is
assigned to each of the data points. We consider three different data sets in
the vicinity of the $f_0(980)$, each of them containing 13 data points (see
Fig:~\ref{fig:errors2}):

\begin{enumerate}

\item
The data set from the levels 2 and 3, using periodic b.c.

\item
The level 2, using both periodic and antiperiodic b.c. As we shall see, the
use of the twisted b.c. enables one to achieve a better accuracy than in set 1.

\item
The levels 2 and 3, periodic b.c., modified input (the parameter
$\eta=0.6$, so the $f_0(980)$ pole is absent). As seen from
Fig.~\ref{fig:errors2}, there is no qualitative change of behavior of the
energy levels in sets 1 and 3. It is important to check, whether the method is
capable to distinguish  between these two cases.

\end{enumerate}

Further, for each set the data $E_i(L)$ shown in Fig.~\ref{fig:errors2} are
fitted by using linear functions in  the potential $V_{ij}$,
\be
V_{ij}=a_{ij}+b_{ij}(s-4M_K^2)\, 
\label{fitv}
\ee
which is used to calculate $E^{\rm (theory)}_i(L)$ by using Eq.~(\ref{eq:det}).
The six parameters of the potential $a_{ij}$ and $b_{ij}$ are assumed to be
free and are determined from the fit to the spectrum. Note that, since the
input potential which was used to produce the data points is also linear in
$s$, the $\chi^2$-value of the best fit is equal to 0.

\begin{figure}
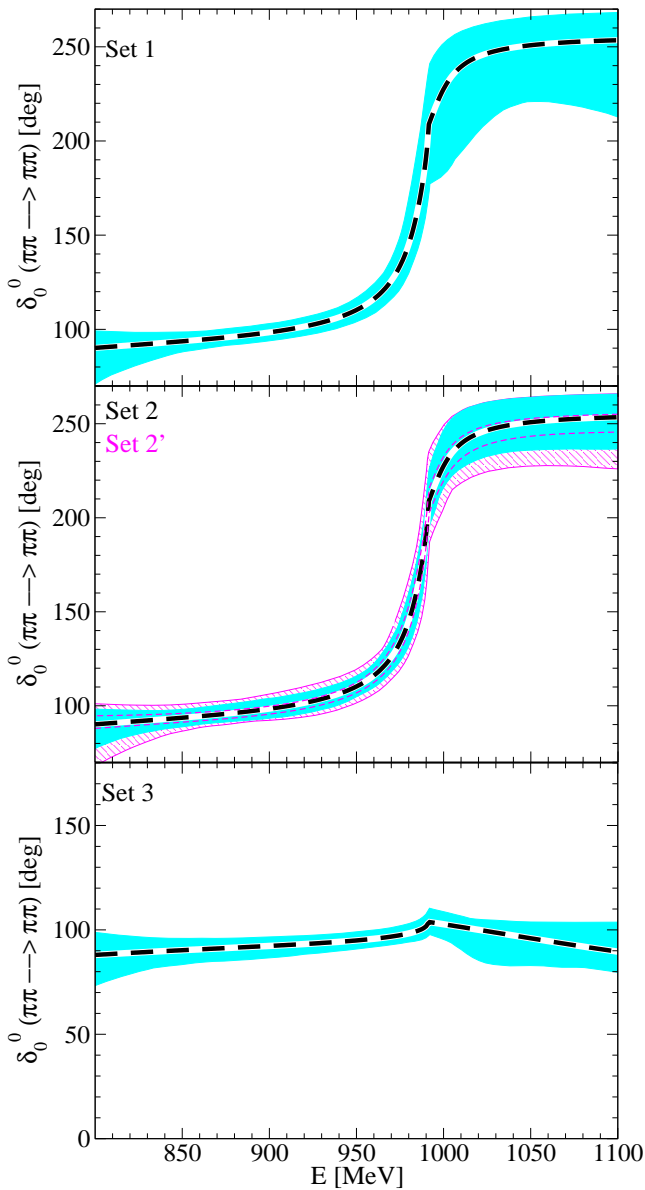

\includegraphics[width=0.45\textwidth]{phase_shift_fitted_w_fo_without_twisted_for_paper_shape.eps}\\

\vspace*{-0.45cm}

\includegraphics[width=0.45\textwidth]{phase_shift_fitted_w_fo_mixed_BC_for_paper_shape.eps}

\vspace*{-0.08cm}

\hspace*{-0.01cm}\includegraphics[width=0.4665\textwidth]{phase_shift_fitted_without_fo_without_twisted_shape.eps}\\

\caption{Extracted phase shifts corresponding to the three  sets shown in
Fig.~\ref{fig:errors2}.  Dashed lines: The calculated phase shifts by using the
approach of  Ref.~\cite{npa}; Filled (cyan) bands: Reconstruction of these phase
shifts.}
\label{fig:fitted_phases}
\end{figure}

\begin{figure}
\includegraphics[width=0.48\textwidth]{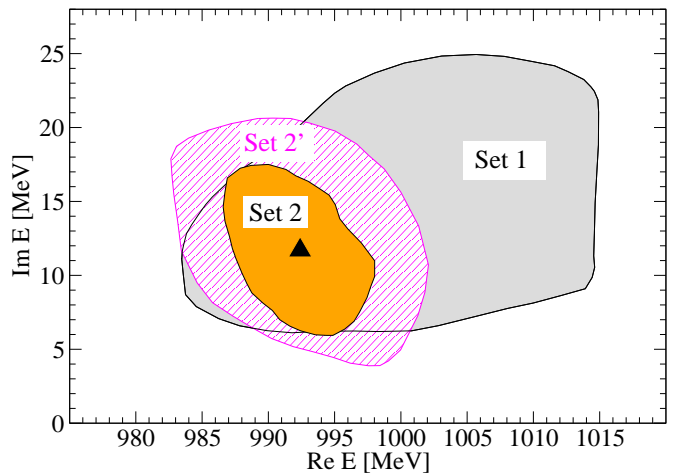}
\caption{Extracted pole positions. 
Filled triangle: original pole position~\cite{npa}; 
Shapes: 
Pole positions reconstructed from the data sets 1, 2, and $2'$, see 
Fig.~\ref{fig:errors2}. For the data set 3, the poles are found 
either on the hidden sheets, or very far from the $K\bar K$ threshold.
}
\label{fig:extracted_poles}
\end{figure}

To obtain the error on the extracted quantities (phase shifts and pole 
positions), random $a_{ij},\,b_{ij}$ are generated within the limits given  by
the parameter errors, that are determined by the previous fit.   Of these
combinations, only those are accepted for which the  resulting $\chi^2$ is
smaller than $\chi_{\rm best}^2+1$.  For each such event, the $\pi\pi$ phase is
evaluated and the pole  position  in the complex energy plane is determined. 
The resulting bands for the phase shifts and extracted  pole positions are
shown in  Figs.~\ref{fig:fitted_phases} and \ref{fig:extracted_poles}. 

From these figures one observes that, for all fits to the synthetic data with
$f_0(980)$ (sets 1,2),  which obey the $\chi_{\rm best}^2+1$ criterion,  the
phase shift shows a resonant behavior and a pole is found in  the complex plane
on the first $K\bar K$ sheet (second sheet for  the $\pi\pi$ channel).  Also,
in all fits to data without $f_0(980)$ (set 3), the phase  shifts show no
resonant behavior, and no pole is found in the  vicinity of the $K\bar K$
threshold --- the poles are in this case  on a hidden sheet far from the
physical region.  Thus, with the method proposed here, it is indeed possible
to  distinguish clearly the presence or absence of a resonance at  the $K\bar
K$ threshold. 

Moreover, as the figures \ref{fig:fitted_phases} and  \ref{fig:extracted_poles}
show, using 13 data points with a 10 MeV error,  the pole position of the
$f_0(980)$ can be determined quite precisely.  As  seen in
Fig.~\ref{fig:extracted_poles}, the data set 2 gives a  more precise
determination of the pole position than the data set 1: it is advantageous to
use data from a lower level but with different boundary conditions, than to use
the data from higher excited levels.

If fewer lattice data are included in the analysis than those of 
Fig.~\ref{fig:errors2}, the six-parameter fit is less constrained.  Then, the
spread in extracted pole positions and phase shifts can increase.  This is
reflected in a blow-up of the bands shown in Fig.~\ref{fig:errors2},  in
regions without data. Then, adding a data point precisely in this region  helps
constrain the fit. In an actual lattice calculation, such  observations could
help to guide the selection of lattice sizes $L$ for which  the levels are
calculated.

\subsection{Statistical scattering of the data (Set $2'$)}
\label{sec:twoprime}
In an actual lattice simulation the finite statistics will not only contribute
to the error bar of a data point but also lead to a displacement of its
centroid. To study this effects, we concentrate on data set 2 and choose a
Gaussian distribution with $\sigma=5$~MeV for the displacement of the centroids.
Using Eq.~(\ref{gauss}) for every data point, 13 new points for Set 2 are
generated and then fitted. The resulting fit leads to a slightly modified pole
position and phase shift. To obtain, however, reliable results, this procedure
has to be repeated multiple times. The multiple generated data sets are called
Set $2'$ in the following. To obtain confidence regions for levels and phase
shifts, the multiple fits to Set $2'$ are analyzed. For example, for the
resulting phase shifts, at every energy $E$ the mean value and standard
deviation for all fits are calculated. The resulting band is indicated with the
dashed lines in Fig.~\ref{fig:fitted_phases}. The same can be made for the
levels themselves as indicated with the dashes lines in Fig.~\ref{fig:errors2}. 

To estimate the combined effect of uncertainty from the previously discussed
$\chi_{\rm best}^2+1$-criterion and displaced centroids, we calculate the
average $\chi_{\rm ave}^2$ from the multiple fits to set $2'$, with the result
$\chi_{\rm ave}^2\sim 1.7$. The overall uncertainty is then estimated from
applying a  $\chi_{\rm best}^2+\chi_{\rm ave}^2+1$-test on set 2, i.e. the data
set without displaced centroids. The resulting bands are indicated with the
hatched areas in Figs.~\ref{fig:errors2} and \ref{fig:fitted_phases}. As
expected, their widths is approximately given by the sum of widths from the
$\chi_{\rm best}^2+1$-criterion [solid (cyan) areas] plus the 1-$\sigma$ region
of the multiple fits to set $2'$ [dashed lines]. In
Fig.~\ref{fig:extracted_poles}, the combined uncertainty for the pole position
is shown [hatched area].  In summary, we observe that the scattering of the
centroids leads to a moderate increase of the uncertainties for levels, phases
and pole positions. With a Gaussian distribution of 5~MeV width for the
displacement of the centroids, the uncertainties induced by 10~MeV error bars
are increased by maximally 50 \%.

\subsection{Systematic uncertainties}
\begin{figure}
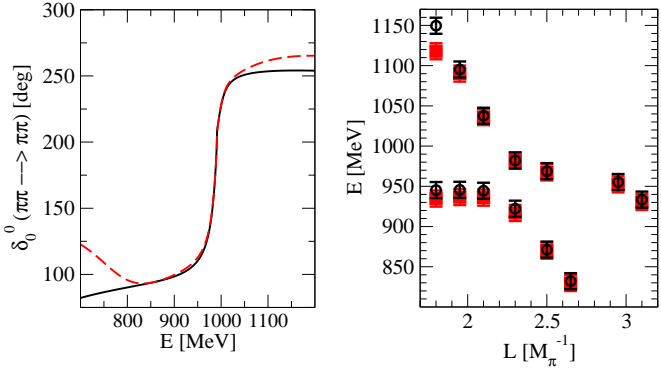

\raisebox{0.2cm}{
\includegraphics[width=0.23\textwidth]{phase_shift_w_fo_SMALL_PERTURBATION.eps}
}
\hspace*{0.05cm}
\includegraphics[width=0.23\textwidth]{levels_w_fo_without_twisted_SQUARED_PERTURBATION.eps}
\caption{The effect of introducing the quadratic term proportional to $c_{ij}$,
on the scattering phase $\delta_0^0$ (left panel), and on the energy spectrum
(right panel). The solid curve and the circles correspond to $c_{ij}=0$. The
dashed line on the left-hand side and the squares on the right-hand side
correspond to $c_{ij}\neq 0$ (small quadratic term in the potential, see
text).}
\label{fig:cij}
\end{figure}

A comprehensive treatment of systematic uncertainties would require to address
the actual lattice action that generates the levels, but that is far beyond of
what can be possibly done in the present framework. What can be studied,
however, is the scenario when the chosen fit potential cannot achieve a perfect
fit of the lattice data. As mentioned above, the choice of the potential of
Eq.~(\ref{fitv})  replicates the lowest-order chiral interaction that has been
used to  generate the synthetic data. To test the more realistic situation,  in
which the lattice data can not be perfectly fitted, apart from the statistical
scattering of the data discussed in the previous section, we introduce a  small
term $\sim c_{ij}\,(s-4M_K^2)^2$ in the potential that is used to  generate the
synthetic data. The constants $c_{ij}$ are chosen to be
$c_{11}=c_{22}=-c,~c_{12}=c_{21}=+c$ where $c=310.8~\mbox{GeV}^{-4}$. The effect
of introducing this term on the phase $\delta_0^0$ and on the energy spectrum
(periodic b.c.) is seen in Fig.~\ref{fig:cij}. The fit to the modified data
points (squares in Fig.~\ref{fig:cij}) and subsequent analysis  are then
performed using the same potential as before, i.e. Eq.~(\ref{fitv}).

\begin{figure}
\includegraphics[width=0.43\textwidth]{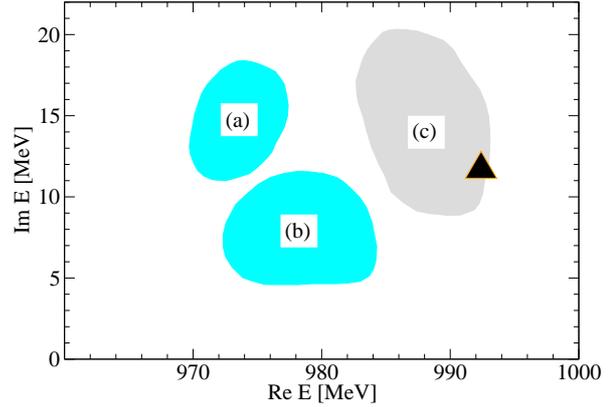}
\caption{Systematic errors. Large triangle: Actual pole position;  (a), (b):
effect of slightly shifted lattice data points (cf. Fig.~\ref{fig:cij}) on the
reconstructed pole position, for the data sets 1 and 2  shown in
Fig.~\ref{fig:errors2};  (c): Cut-off-dependence ($q_{\rm max}=755$~MeV instead
of $904$ MeV  as used in the generation of the data).}
\label{fig:extracted_poles_errors}
\end{figure}

The resulting $\chi^2_{\rm best}$ does not vanish any more, and  the extracted
pole positions and phases are systematically  shifted as shown in
Fig.~\ref{fig:extracted_poles_errors}, case (a) and (b) (the latter correspond
to data sets 1 and 2 in Fig.~\ref{fig:errors2}. In addition, to make these
areas distinguishable, we have assigned an error  $\Delta E=5~\mbox{MeV}$ in
case (a), instead of 10~MeV.).
This kind of systematic error is inherently tied to the assumption  made on the
functional form of the potential (cf. Eq.~(\ref{fitv})):  one could, of course,
introduce higher order terms in the potential  to extract phase shifts and pole
positions. Then, $\chi^2_{\rm best}$ would  be close to zero again, but the fit
would be much less constrained  given the higher number of free parameters. The
spread of pole positions  and extracted phases would then immediately increase
as  compared to the present results. Still, it would
be worth to quantify the effects of the higher order terms on the
accuracy of these determinations in future studies.

Apart from the discussed higher order terms for the potential $V$,  another
source of uncertainty is given by the cutoff $q_{\rm max}$ that  is used to
determine phase shifts and pole positions once the $V_{ij}$  are fitted: the
term $\tilde G$ of Eq.~(\ref{eq:det}) depends  on this $q_{\rm max}$. To test
the dependence, we have extracted the  pole positions with $q_{\rm max}=755$
MeV instead of using the standard  value of $904$ MeV that has been used to
generate the synthetic data.  The result is shown as case (c) in 
Fig.~\ref{fig:extracted_poles_errors}. Obviously, the functional form of
Eq.~(\ref{fitv}) in $V_{ij}$ is flexible enough  to compensate for this large
change in $q_{\rm max}$ and as a  result the pole position is shifted by just a
few MeV.  Systematic effects from the cut-off dependence are, thus, very small.

We have made another test which can be of use if one could make a rough 
estimate of the size of the systematic uncertainties of the lattice 
results.  Systematic uncertainties would most likely produce deviations 
of the  energy levels in the same direction.  We have assumed a constant 
shift of the energy levels, increasing them by 5~MeV, and we have found 
a decrease of about 5\% in the phase shifts, and viceversa.

To summarize, a careful examination of all kind of errors has been carried out
in this subsection, illustrating the limitations of the proposed method. Given
only a few lattice  data with large error, UCHPT in finite volume  is certainly
a useful tool to make  reliable statements on the presence or absence of
resonances,  and even to give a good first estimate on phases and pole
positions.

\subsection{The $a_0(980)$}
\label{subsec:a0}

\begin{figure}
\includegraphics[width=0.49\textwidth]{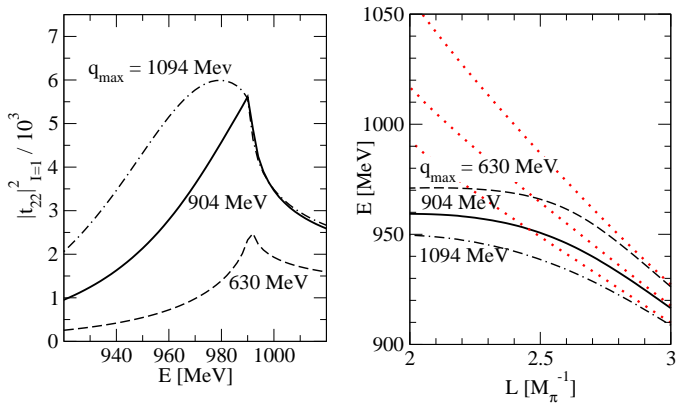}
\caption{Left: lineshapes in $\pi\eta\to\pi\eta$.  Note that the $a_0$ shows a
true resonance pole for  $q_{\rm max}=1094$ MeV, while for $q_{\rm max}=904$
and $630$ MeV,  it is rather a cusp at the $K\bar K$ threshold than a
resonance  (in this case, the pole is on a hidden sheet); Right: Level 1
for three different  cutoffs $q_{\rm max}=1094,\, 904, \,630,$ MeV, with
periodic (solid, dashed, dash-dotted) and antiperiodic (dotted) b.c.}
\label{fig:aospectrum}
\end{figure}

The situation concerning the $a_0(980)$ in UCHPT is intriguing.  For example, in
Ref.~\cite{npa}, the $a_0(980)$ is generated from the lowest order chiral
Lagrangian in the $\pi\eta$ and $K\bar K$ channels. At present, there exists no
consensus about its nature and exact location. For example, changing the cutoff
$q_{\rm max}$ in the approach of Ref.~\cite{npa}, one may easily  move the pole
below/above the $K\bar K$ threshold. Namely, for $q_{\rm max}=1094$ MeV, the
pole of the $a_0$ is below the  $K\bar K$ threshold and one observes the full
resonance shape (see the left panel of Fig.~\ref{fig:aospectrum}). For lower
cut-offs, say $q_{\rm max}=904$~MeV and $q_{\rm max}=630$~MeV, the pole
disappears from the second  Riemann sheet and instead of a resonance shape, a
pronounced threshold cusp structure  is visible. This was already observed in
Refs.~\cite{npa,Oller:1998hw}. 

This is an interesting situation, because there are several examples in the 
literature where it is not clear if an observed structure is a resonance  or
rather a cusp (example: a disputed pentaquark in  $\gamma n\to\eta
N$~\cite{Kuznetsov:2007gr,Shklyar:2006xw,Doring:2009qr,Anisovich:2008wd}).
Using UCHPT in a finite volume, one may predict the dependence of the
energy levels on the cutoff parameters and see, whether there is a clear-cut
distinction between the situations with a resonance or a cusp.

In order to show how the different scenarios influence the energy levels in the
box, we carry out the calculation and show the results in
Fig.~\ref{fig:aospectrum}. In the left panel of this figure we show the
lineshapes in $\pi\eta\to\pi\eta$ for three different cutoffs. In the right
panel, we display the energy level 1 for the same values of the
cutoff, both for periodic and antiperiodic b.c. As one observes from this plot,
the level is quite sensitive to the value of the cutoff. Consequently, it is
expected that the extraction of the pole position from future lattice data  can
be performed accurately, resulting in a clear-cut resolution of the
resonance/cusp scenario.

\subsection{The $f_0(600)$}
\label{sec:sigma}

\begin{figure*}
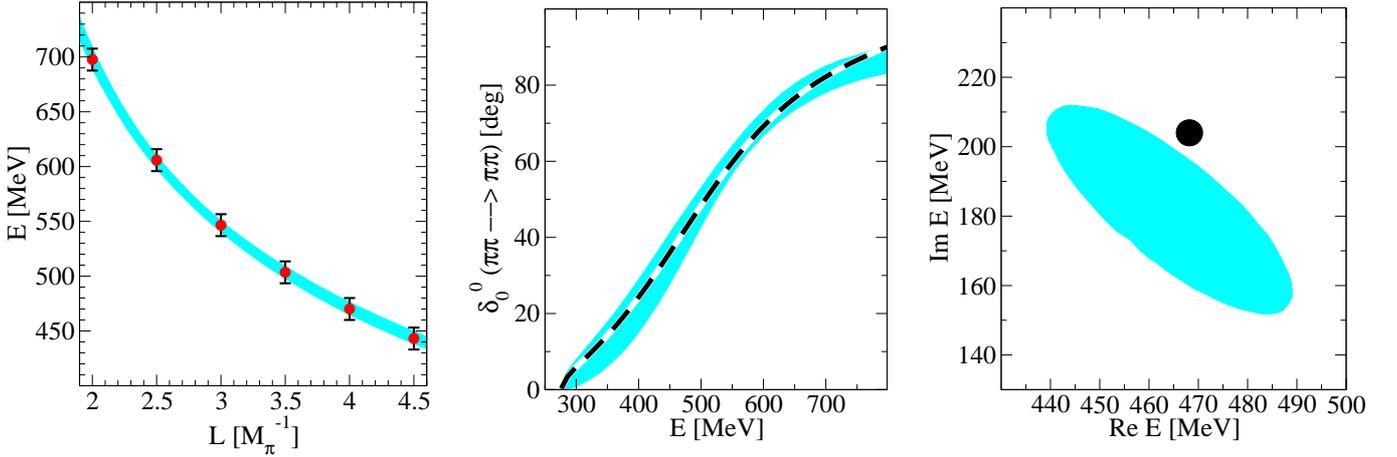

\includegraphics[width=0.31\textwidth]{levels_sigma600_shape.eps}
\hspace*{.2cm}
\raisebox{0.2cm}{
\includegraphics[width=0.31\textwidth]{phase_shift_fitted_sigma_600_shape.eps}
}
\hspace*{.2cm}
\raisebox{0.2cm}{\includegraphics[width=0.32\textwidth]{pole_spread_sigma600.eps}
}
\caption{Left: Synthetic lattice data (with errors) for the energy level 1;
Center: The extracted phase shift vs the original $\delta_0^0$ phase shift
(dashed line); Right: The spread in the pole positions. The filled circle shows
the original pole position.}
\label{fig:sigma}
\end{figure*}

It comes rather as a surprise that the procedure seems to work even in case  of
the $f_0(600)$. In order to do this, we have resorted to the 1-channel L\"uscher
formalism, because the energy is well below the $K\bar K$ threshold. Only the
level 1 with periodic b.c. was fitted (see Fig.~\ref{fig:sigma}, left panel).
The fit resulted in the phase shifts shown in the center panel and in the pole
positions displayed in the right panel of the same figure. As seen from the
figure, the spread in pole positions is larger than in case of $f_0(980)$, but,
in our opinion, it is remarkable that one is able to address the question of the
extraction of the $f_0(600)$ on the lattice at all. Note also that, as seen from
Fig.~\ref{fig:sigma}, the small effect of the sub-threshold $K\bar K$ channel 
is visible in the extracted phase for the highest energies above 600~MeV.  Also,
for the same reason, there is an off-set between the true pole position (dark
circle, right-hand panel) and the extracted ones (large shaded area). To verify
this, we  have generated pseudo data using only the $\pi\pi $-channel in the
hadronic model~\cite{npa}. Then, extracting pole positions with the one-channel
formalism, the resulting ellipse should have the true pole position (of the
$\pi\pi$-reduced hadronic model) in its center. This is indeed the case as has
been checked.

The $f_0(600)$ (or $\sigma$) has been predicted in most UCHPT approaches (see,
e.g.,  Refs.~\cite{npa,Locher:1997gr,Oller:1998hw}). Recently,  its existence
has been rigorously proved by using ChPT combined with Roy
equations~\cite{Caprini:2005zr}. It would be intriguing, if the already existing
lattice studies of the lowest scalar resonance in 
QCD~\cite{Mathur:2006bs,Kunihiro:2003yj,Prelovsek:2010gm} could be  supplemented
by the novel method of the analysis based on UCHPT,  in order to facilitate the
extraction of the pole position in the complex plane.

\section{Conclusions}
\label{sec:concl}

In this paper, we discuss the extraction of the parameters of the $f_0(600)$,
$f_0(980)$ and $a_0(980)$ resonances from lattice data. In order to facilitate
this extraction, we use UCHPT in a finite volume in the fit. Fitting  synthetic
data, we have demonstrated that the approach works, and the pole position for
the above resonances can indeed be extracted by analyzing the   volume-dependent
energy spectrum in the vicinity of the resonance energy, provided sufficiently
many volumes are simulated. The different sources of errors are analyzed in
detail. The key point is that the use of the phenomenological input from UCHPT
stabilizes the fit.  It is, thus, very challenging to apply these results to
present and forthcoming lattice data in order to extract valuable information
concerning hadronic resonances. From the theoretical point of view one should
also keep in mind that in field theory there are parts of the kernel of the
Bethe-Salpeter equation which are volume dependent, although exponentially
suppressed. Quantitative studies of these terms and the extent of the large $L$
suppression would be a very good complement to the work we have carried out
here.

\section*{Acknowledgments}
The authors would like to thank C. Alexandrou,
Ch. Lang,  M. Peardon, S. Prelovsek, M. Savage, and C. Urbach for
discussions. This work is partly supported by DGICYT contracts  FIS2006-03438,
FPA2007-62777, the Generalitat Valenciana in the program Prometeo and  the EU
Integrated Infrastructure Initiative Hadron Physics Project  under Grant
Agreement n.227431. We also acknowledge the support by DFG (SFB/TR 16,
``Subnuclear Structure of Matter''),  by the Helmholtz Association through funds
provided to the virtual institute ``Spin and strong QCD'' (VH-VI-231)  and by
COSY FFE under contract 41821485 (COSY 106).  A.R. acknowledges  support of the
Georgia National Science Foundation (Grant \#GNSF/ST08/4-401).

\appendix
\section{Quantitative comparison to the  L\"uscher approach}
\label{sec:appendix}

As we have seen in Sec.~\ref{sec:relation}, the connection of our approach 
with the L\"uscher approach comes through the replacement $\tilde G_j(E) \to
\tilde G_{jA}(E)$, where
\ba
\tilde G_{jA}(E)-G_j(E)&=&-\frac{1}{4\pi^{3/2}EL}\,{\cal Z}_{00}(1;\hat p^2)
\non&+&
\frac{i}{16\pi E^2}\,\lambda^{1/2}(E^2,m_1^2,m_2^2)\, .
\label{rela1}
\ea
It is straightforward to see that, substituting $G$ in Eq.~(\ref{bse}) by
$\tilde G_{jA}$ of Eq.~(\ref{rela1}), one arrives at the two-channel L\"uscher
equation from  Ref.~\cite{akaki}, which is written in terms of the $K$-matrix.

\begin{figure}
\includegraphics[width=0.48\textwidth]{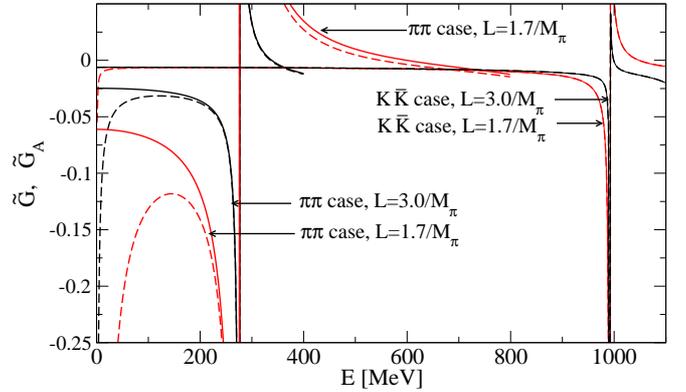}
\caption{Comparison of $\tilde G_j(E)$ of Eq.~(\ref{tildegs}) [solid lines]
and  $\tilde G_{jA}(E)$ of Eq.~(\ref{rela1}) [dashed lines],  for the $\pi\pi$
and $K\bar K$ channels.  Two cases for $L=1.7/M_\pi$ and $L=3.0/M_\pi$ are
shown.}
\label{fig:blubb}
\end{figure}

Next we discuss the differences of the functions $\tilde G_j(E)$ of
Eq.~(\ref{tildegs})  and  $\tilde G_{jA}(E)$. In Fig.~\ref{fig:blubb}  we show
the results for two different  values of $L$ for the $\pi\pi$ and the $K\bar K$
channels. 
We can see that for the $\pi\pi$ channel the differences for 
$L=1.7\,M_\pi^{-1}$ are rather large  for energies below 250~MeV,  which is
already below the $\pi\pi$ threshold. Even above threshold, the differences are
sizable. These differences  are smaller for a bigger $L$, $L=3.0\,M_\pi^{-1}$,
as expected for  exponentially suppressed terms. In both cases the function 
$\tilde G_{jA}(E)$ has a pole at $E=0$, while $\tilde G_j(E)$ is finite there.
For the case of the $K\bar K$ channel, the trend is similar to the $\pi\pi$
channel, but the region where $\tilde G_{jA}(E)$ and $\tilde G_j(E)$ are 
practically equal is larger than in the $\pi\pi$ case.

\begin{figure*}
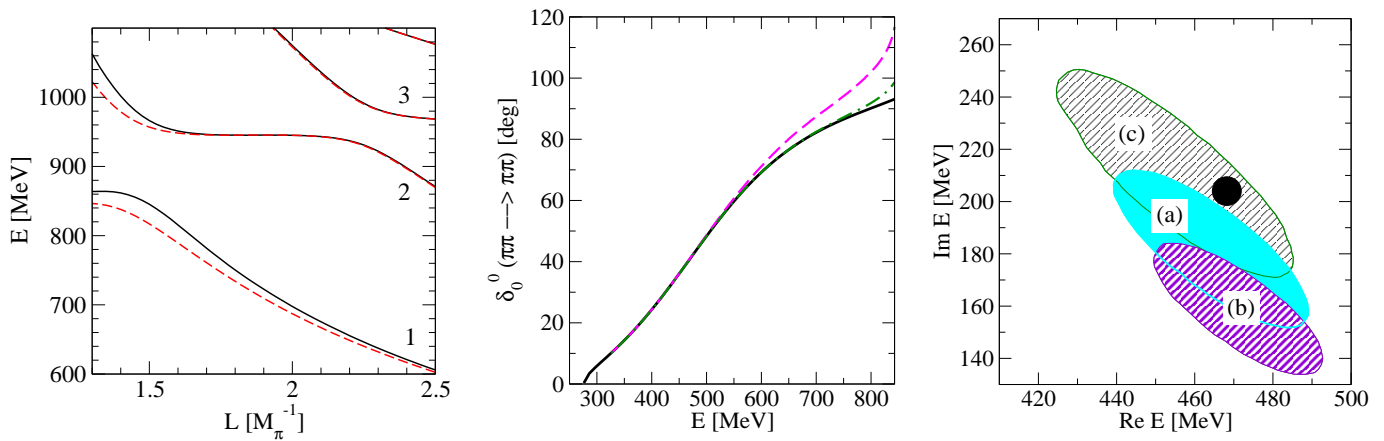

\includegraphics[width=0.323\textwidth]{compare_with_levels_from_GA.eps}
\hspace*{.2cm}
\raisebox{0.1cm}{
\includegraphics[width=0.3\textwidth]{phase_shift_fitted_sigma_600_Zoo_effect.eps}
}
\hspace*{.2cm}
\raisebox{0.1cm}{\includegraphics[width=0.32\textwidth]{pole_spread_sigma600_test_Z00_shapes.eps}
}
\caption{Left panel:  Levels obtained with $\tilde G_j(E)$ of
Eq.~(\ref{tildegs})  [solid lines, identical to
Fig.~\ref{fig:levels_nonsmooth}] and with  $\tilde G_{jA}(E)$  of
Eq.~(\ref{rela1}) [dashed lines]. Center panel, dashed and dash-dotted lines: 
Pseudophases (one-channel formalism) obtained from the levels shown in the left
panel [see text], using Eq.~(\ref{eq:pre_luescher}). Solid line: Original
phase. Right panel: Pole positions for $f_0(600)$, extracted by using the
method of section~\ref{sec:sigma}. For cases (a), (b), (c), see text.}
\label{fig:appendix2}
\end{figure*}

It is interesting to see,  how big the effect of this difference is on the
energy levels in the box. This is seen in Fig.~\ref{fig:appendix2}, left panel.
As we see from Fig.~\ref{fig:appendix2}, left panel, if  $L\geq
1.6\,M_\pi^{-1}$, the corrections to the energy levels 2,3,\ldots are
completely negligible. Therefore, we concentrate on level 1 where the
difference between the solid and dashed curves can not be neglected.  We
calculate the $\pi\pi$ pseudophases, see Fig.~\ref{fig:appendix2},  central
panel, and the position of the poles of the $f_0(600)$ corresponding to these
two curves, see Fig.~\ref{fig:appendix2}, right panel~\footnote{For small
volumes, formally exponentially small terms $\exp(-n\,M_\pi\,L)$ are no longer
suppressed and would have to be evaluated, too.}.      

As one sees from Fig.~\ref{fig:appendix2}, the pseudophases are fairly the same
until 500~MeV, from where they start diverging.  At $L=2\, M_\pi^{-1}$ that
corresponds to $E\simeq 700~\mbox{MeV}$, the difference between two
pseudophases is around 5~degrees. If the energy grows (the volume decreases),
the difference between the two pseudophases becomes, as expected, bigger.
 
In Fig.~\ref{fig:appendix2}, right panel, we display the shift of the pole
positions due to the effect of using the relativistic propagator. The same
method as in section~\ref{sec:sigma} was used to produce the ellipses.  Case
(a) shows the pole positions extracted from levels generated using the
relativistic propagator (identical case to the one shown in
Fig.~\ref{fig:sigma}). In case (b) the pole positions have been extracted from
the level generated with $\tilde G_{jA}(E)$  of Eq.~(\ref{rela1}), i.e. from
the dashed curve shown in the left panel. As we see, the real part
of the pole undergoes a shift of order of 10~MeV, whereas the width (twice  the
imaginary part of the pole position) changes by approx. 40~MeV.

As we see, the uncertainty of the method which stems from the error  of 10~MeV
in energy attached to each data point, is still larger than the change induced
by the use of the relativistic propagators.  Should one aim at an accuracy of
better than 40~MeV in the width,  first, the errors in the lattice data should
improve beyond 10~MeV and, second, from our study it follows that the standard
L\"uscher approach is not accurate enough in this case.

So far, we were concerned how a shift in the level translates into a shift of
the pseudophase and pole position; one may also ask what happens if $\tilde
G_{jA}(E)$ of Eq.~(\ref{rela1}) is used to {\it extract} the pole position from
a given level, instead of using the relativistic propagator $\tilde G_j(E)$ of
Eq.~(\ref{tildegs}) as done before. The effect is shown as case (c) in the
right-hand panel of Fig.~\ref{fig:appendix2}. Compared to case (a), one 
observes a systematic shift of the results, which is of similar size as the
previously discussed case (b).

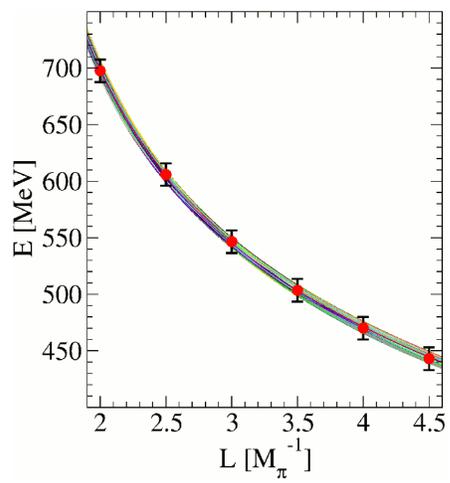
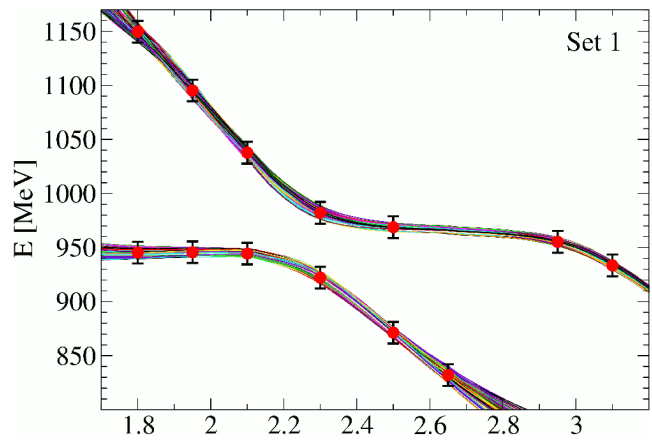
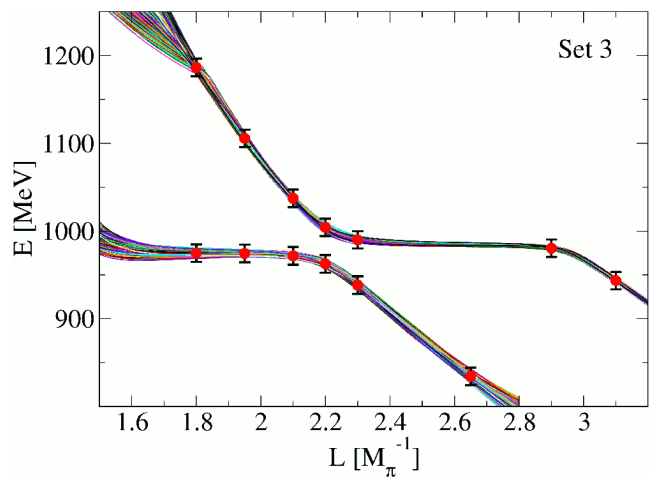
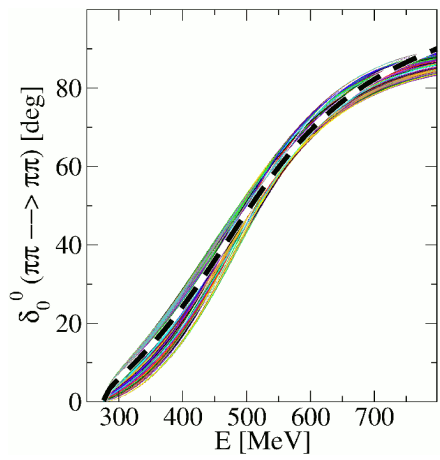
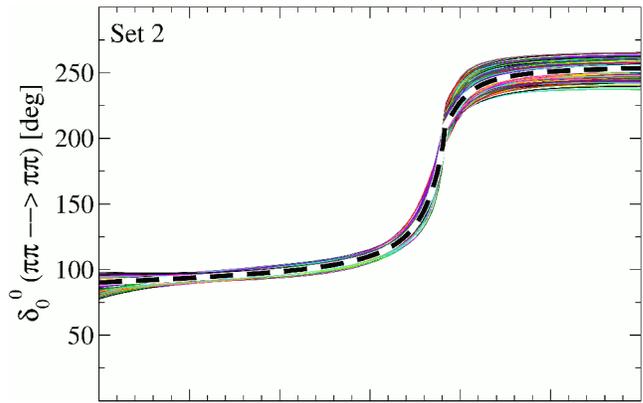
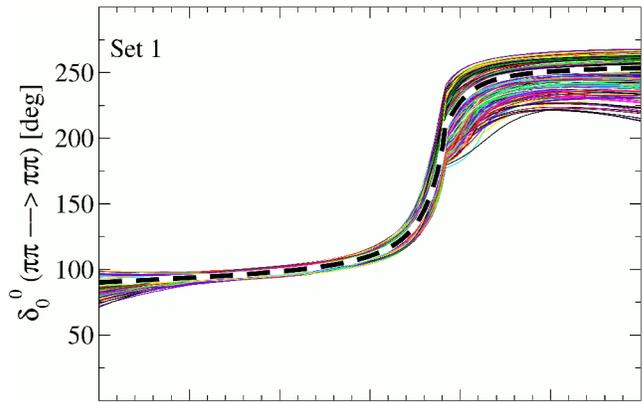
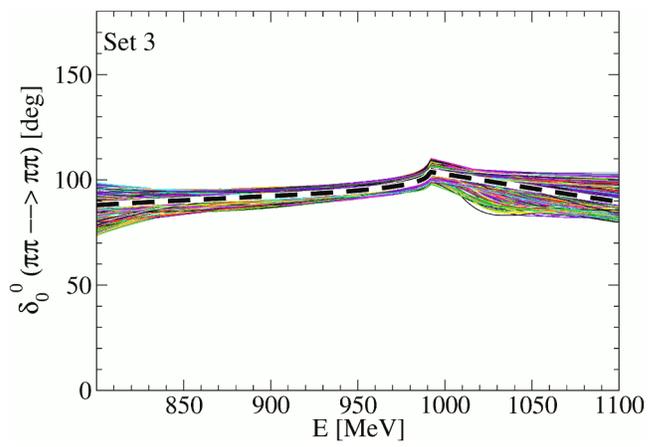
\end{document}